\documentclass[%
reprint,
superscriptaddress,
 amsmath,amssymb,
 aps,
]{revtex4-1}

\usepackage{graphicx}% Include figure files
\usepackage{dcolumn}% Align table columns on decimal point
\usepackage{bm}% bold math

\usepackage{braket}%Dirac notation for states

\usepackage{xcolor}
\usepackage[normalem]{ulem}

\newcommand{\rev}[1]{\textcolor{black}{#1}}
\usepackage[pagebackref=false]{hyperref}

\begin{document}

\author{L.~Banszerus}
\thanks{These authors contributed equally to this work}
\affiliation{JARA-FIT and 2nd Institute of Physics, RWTH Aachen University, 52074 Aachen, Germany,~EU}%
\affiliation{Peter Gr\"unberg Institute  (PGI-9), Forschungszentrum J\"ulich, 52425 J\"ulich,~Germany,~EU}

\author{K.~Hecker}
\thanks{These authors contributed equally to this work}
\affiliation{JARA-FIT and 2nd Institute of Physics, RWTH Aachen University, 52074 Aachen, Germany,~EU}%
\affiliation{Peter Gr\"unberg Institute  (PGI-9), Forschungszentrum J\"ulich, 52425 J\"ulich,~Germany,~EU}

\author{S.~M\"oller}
\affiliation{JARA-FIT and 2nd Institute of Physics, RWTH Aachen University, 52074 Aachen, Germany,~EU}%
\affiliation{Peter Gr\"unberg Institute  (PGI-9), Forschungszentrum J\"ulich, 52425 J\"ulich,~Germany,~EU}

\author{E. Icking}
\affiliation{JARA-FIT and 2nd Institute of Physics, RWTH Aachen University, 52074 Aachen, Germany,~EU}%
\affiliation{Peter Gr\"unberg Institute  (PGI-9), Forschungszentrum J\"ulich, 52425 J\"ulich,~Germany,~EU}

\author{K.~Watanabe}
\affiliation{Research Center for Functional Materials, 
National Institute for Materials Science, 1-1 Namiki, Tsukuba 305-0044, Japan
}
\author{T.~Taniguchi}
\affiliation{ 
International Center for Materials Nanoarchitectonics, 
National Institute for Materials Science,  1-1 Namiki, Tsukuba 305-0044, Japan
}

\author{C.~Volk}
\author{C.~Stampfer}
\affiliation{JARA-FIT and 2nd Institute of Physics, RWTH Aachen University, 52074 Aachen, Germany,~EU}%
\affiliation{Peter Gr\"unberg Institute  (PGI-9), Forschungszentrum J\"ulich, 52425 J\"ulich,~Germany,~EU}%

\title{Spin relaxation in a single-electron graphene quantum dot}
\date{\today}% It is always \today, today,


\begin{abstract}
\end{abstract}

\maketitle

\textbf{The relaxation time of a single-electron spin is an important parameter for solid-state spin qubits, as it directly limits the lifetime of the encoded information. Thanks to the low spin-orbit interaction and low hyperfine coupling, graphene and bilayer graphene (BLG) have long been considered promising platforms for spin qubits~\cite{Trauzettel2007Mar}. Only recently, it has become possible to control single-electrons in BLG quantum dots (QDs) and to understand their spin-valley texture~\cite{Eich2018Jul,Kurzmann2019Jul,Banszerus2021Sep}, while the relaxation dynamics have remained mostly unexplored~\cite{Volk2013Apr,Banszerus2021Feb}. Here, we report spin relaxation times ($T_1$) of single-electron states in BLG QDs. Using pulsed-gate spectroscopy, we extract relaxation times exceeding 200~$\mu$s at a magnetic field of 1.9~T. The $T_1$ values show a strong dependence on the spin splitting, promising even longer $T_1$ at lower magnetic fields, where our measurements are limited by the signal-to-noise ratio. The relaxation times are more than two orders of magnitude larger than those previously reported for carbon-based QDs, \rev{suggesting that graphene is a potentially promising host material for scalable spin qubits.}}

The concept proposed by Loss and DiVincenzo to encode quantum information in spin states of  QDs~\cite{Loss1998Jan} has laid the foundation of spin-based solid state quantum computation. Spin qubits have been realized in III-V semiconductors~\cite{Petta2005Sep,Nowack2011Sep,Shulman2012Apr}, as well as in silicon~\cite{Veldhorst2015Oct,Zajac2018Jan,He2019Jul,Xue2022Jan} and germanium~\cite{Hendrickx2021Mar}. The lifetime of the information encoded in such qubits is ultimately limited by the spin relaxation time, $T_1$. This relaxation time can be estimated via transient current spectroscopy, where the excited spin state of the QD is occupied with the help of high-frequency voltage pulses, applied to one of the gates of the QD~\cite{Fujisawa2002Sep,Hanson2003Nov,Fujisawa2002Sep,Volk2013Apr,Banszerus2021Feb}.
\begin{figure*}[]
\centering
\includegraphics[draft=false,keepaspectratio=true,clip,width=0.98\linewidth]{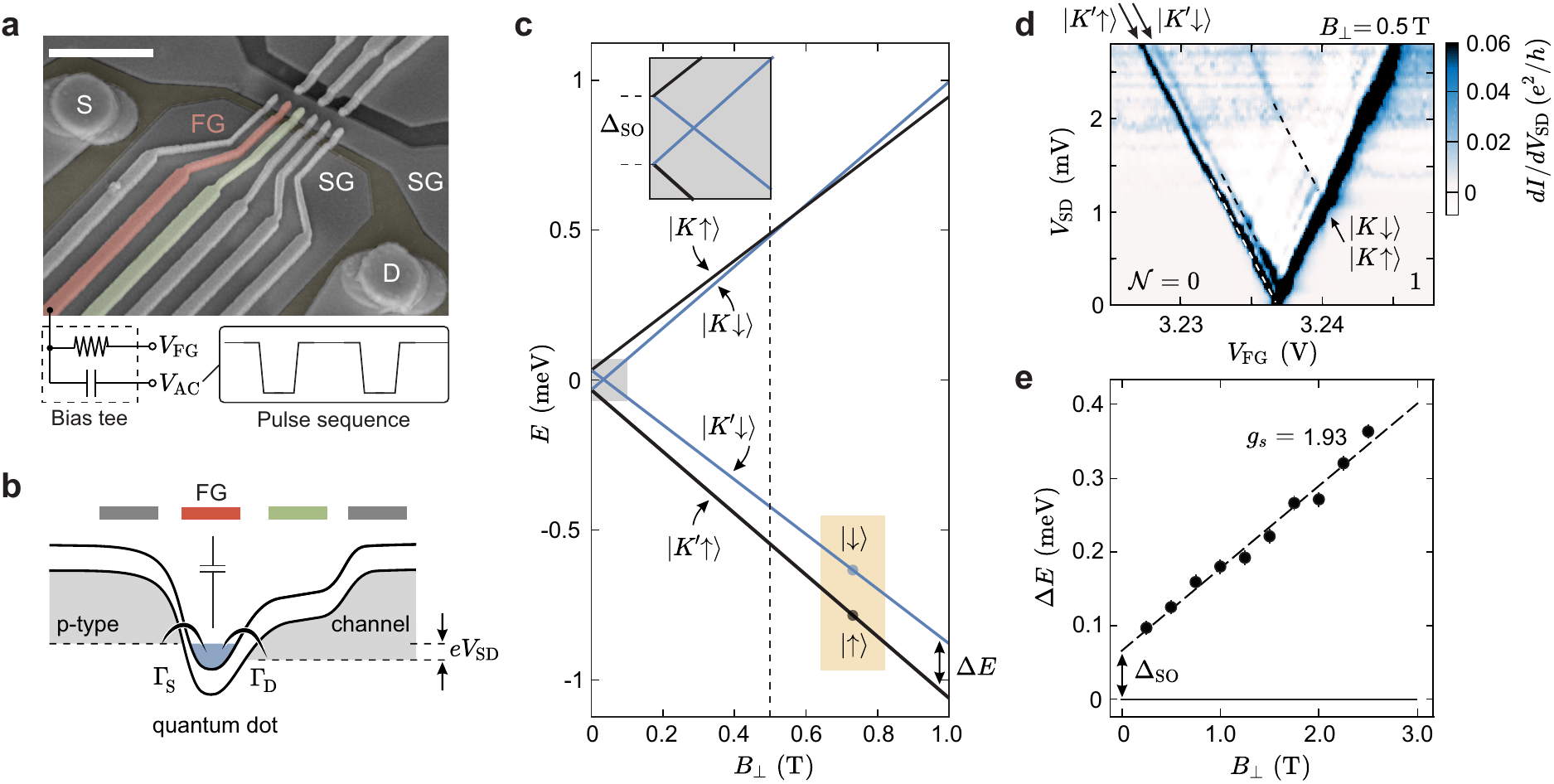}
\caption[Fig01]{\textbf{Device schematics and single-particle spectrum.}
\textbf{a} False-color scanning electron microscopy image of the gate layout. The SGs define a narrow conducting channel connecting source and drain, while the FGs across the channel are used to form a QD. Bias tees connected to the FGs allow the application of AC pulses ($V_\mathrm{AC}$) and DC voltages ($V_\mathrm{FG}$) to the same gate. The scale bar corresponds to 1~$\mu$m.  
\textbf{b} Band schematic along the channel. One FG~(red) is tuned to form a QD, while the tunnel coupling to the right lead, $\Gamma_\mathrm{D}$, is controlled using a neighboring FG~(green). 
\textbf{c} Single particle spectrum of a BLG QD as function of a out-of-plane magnetic field $B_\perp$. Inset: At $B_\perp = 0$~T, the spin-orbit interaction splits the four states of the first orbital into Kramer's pairs with spin-orbit gap $\Delta_\mathrm{SO}$. 
\textbf{d} Finite bias spectroscopy measurement of the single particle spectrum recorded at $B_\perp=0.5~$T (see dashed line in c). Dashed lines highlight the four single particle states.
\textbf{e} Measured energy splitting $\Delta E$ of the two $K'$-states, $\ket{\uparrow}$ and $\ket{\downarrow}$, as a function of $B_\perp$. 
}
\label{f1}
\end{figure*}
In single- and two-electron QDs in GaAs for example, $T_1$ times up to 200~$\mu$s have been reported~\cite{Fujisawa2002Sep,Hanson2003Nov}. Group IV elements such as silicon, germanium and carbon are particularly interesting hosts for realizing spin qubits, \rev{thanks to their low nuclear spin densities and the abundance of nuclear spin free isotopes.} While $T_1$ times of up to one second have been reported for silicon QDs with small spin splittings~\cite{Yang2013Jun}, $T_1$ times of about 10~$\mu$s have been found in carbon nanotube QDs at low magnetic fields~\cite{Churchill2009Apr,Laird2013Aug}. The latter is most likely limited by the curvature-induced spin-orbit interaction in nanotubes on the order of $\Delta_\text{SO}\approx 1$~meV~\cite{Laird2013Aug}. In contrast, flat graphene and BLG exhibit both low hyperfine coupling and small Kane-Mele type spin-orbit interaction on the order of 40-80~$\mu$eV~\cite{Min2006Oct,Huertas-Hernando2006Oct,Konschuh2012Mar,Banszerus2021Sep,Kurzmann2021Mar}, promising long spin lifetimes. \rev{Early devices were based on etched QDs in single-layer graphene, where edge disorder prevented control over the charge occupation of the QDs~\cite{Stampfer2008Aug,Stampfer2009Feb,Guttinger2012Nov}, imposing currently a major roadblock for single-layer graphene based qubits. In contrast,}
BLG is particularly suitable for realizing highly tunable QDs
~\cite{Eich2018Jul,Banszerus2018Aug}, and important steps towards the realization of spin qubits  have already 
been achieved -- such as the implementation of charge detection~\cite{Kurzmann2019Aug,Banszerus2021Mar}, the investigation of the electron-hole crossover~\cite{Banszerus2020Oct} and the measurement of the spin-orbit gap in BLG~\cite{Banszerus2020May,Banszerus2021Sep,Kurzmann2021Mar}.
However, electrical measurements of the  spin relaxation time have remained elusive \rev{in both, single-layer graphene and BLG} until now. In this letter, we report on the measurement of $T_1$ \rev{times} in a single-electron BLG QD. \rev{Our measurements confirm that the relaxation time is sufficiently long to potentially operate a spin qubit, making graphene an interesting host material for bench-marking spin qubits.}

The device consists of a BLG flake encapsulated in hexagonal boron nitride~(hBN) placed on a global graphite back gate (BG), with two layers of metallic top gates. Fig.~\ref{f1}a shows a scanning electron microscopy image of the gate structure of the device~(see methods for details)~\cite{Banszerus2020Oct}. To form a QD, we
use the BG and split gates (SG) to form a p-type channel connecting source \rev{(S)} and drain \rev{(D)}. The potential along the channel can be controlled using a set of finger gates~(FGs) and a QD is formed by locally overcompensating the potential set by the BG using one of the FGs (see red FG in Fig.~\ref{f1}a,b), forming a p-n-p junction, where an n-type QD is tunnel coupled to the p-type reservoirs. The electron occupation of the QD can be controlled down to the last electron using the FG potential $V_\mathrm{FG}$ \rev{(see supplementary Fig.~S1 and S2 for details)}. The tunnel coupling between the QD and the channel can be tuned using adjacent
FGs (e.g. green FG in Fig. 1a,b), allowing also to realize configurations with strongly asymmetric tunnel barriers, as illustrated in the schematic of Fig. 1b. \rev{All other FG potentials are kept on ground.}

Fig. ~\ref{f1}c shows the energy dispersion of the first orbital state of the QD as a function of an external out-of-plane magnetic field. 
In BLG, each single-particle orbital is composed of four states, because of the spin and valley degrees of freedom. In contrast to silicon, the valley states in BLG are associated with topological out-of-plane magnetic moments, which originate from the finite Berry curvature close to the $K$-points and has opposite sign for the 
$K$ and $K'$-valley~\cite{McCann2013Apr}.
At zero magnetic field, the Kane-Mele type spin-orbit interaction~\cite{Kane2005Nov} splits the four degenerate states into two Kramer's pairs $(\ket{K\uparrow},\ket{K'\downarrow})$ and $(\ket{K\downarrow}, \ket{K'\uparrow})$ with an energy gap $\Delta_\mathrm{SO}$~\cite{Banszerus2021Sep,Kurzmann2021Mar} (see inset of  Fig.~\ref{f1}c). An out-of-plane magnetic field, $B_\perp$, lifts the degeneracy and each state shifts in energy according to the spin and valley Zeeman effect as $E(B_\perp) = \frac{1}{2} (\pm g_\text{s} \pm g_\nu) \mu_\text{B} B_{\perp}$,
with the Bohr magneton $\mu_\text{B}$, the spin $g$-factor $g_\text{s}=2$ and the valley $g$-factor $g_\nu$.
Considering typical values of $\Delta_\mathrm{SO} \approx 65~\mu$eV and $g_\nu \approx 30$~\cite{Banszerus2021Sep}, valley polarization of the two lowest energy states is achieved already at about 50~mT. In this regime, the system can be treated as an effective two-level spin system with the ground state $\ket{K'\uparrow}$ and excited state $\ket{K'\downarrow}$ which are split by $\Delta E (B_\perp) = \Delta_\mathrm{SO}+g_\text{s} \mu_\text{B} B_\perp$.

\begin{figure*}[]
\centering
\includegraphics[draft=false,keepaspectratio=true,clip,width=0.95\linewidth]{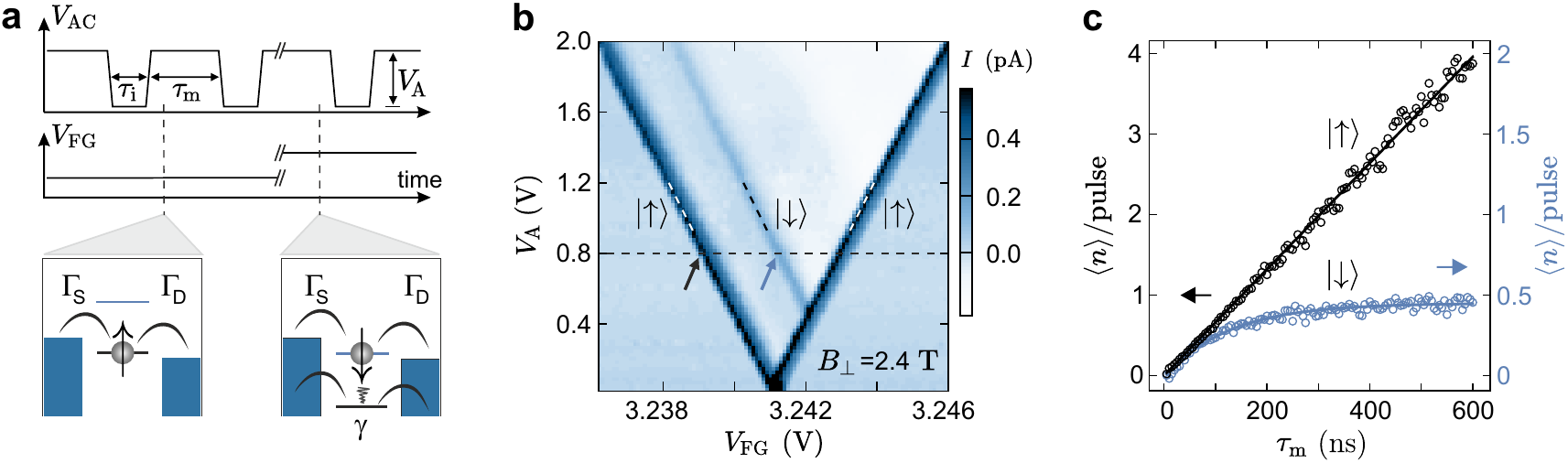}
\caption[Fig02]{\textbf{Transient current spectroscopy.}
\rev{\textbf{a} The schematic depicts a square pulse with amplitude $V_\text{A}$ and pulse widths $\tau_\text{i}$ and $\tau_\text{m}$. Bottom: possible processes if the GS (left) or ES (right) reside in the bias window, which depends on the DC gate voltage, $V_\mathrm{FG}$. 
\textbf{b} Current through the QD as a function of the $V_\mathrm{FG}$ and the pulse amplitude $V_\mathrm{A}$ at the transition from $\mathcal{N}=0 \rightarrow 1$ electrons ($V_\mathrm{SD} = 80~\mu$V, $f=2.5~$MHz, $\tau_\mathrm{m}=\tau_\mathrm{i}$, $B_\perp = 2.4$~T). At low $V_\mathrm{A}$, only $\ket{\uparrow}$-transport is visible during $\tau_\mathrm{i}$ and $\tau_\mathrm{m}$. At $V_\mathrm{A} \approx 0.5~$V, i.e. the pulse excitation exceeding the level splitting, a transient current via $\ket{\downarrow}$ sets in.
\textbf{c} Average number of electrons $\langle n \rangle$ per pulse cycle ($\langle n \rangle/\mathrm{pulse} = I(\tau_\mathrm{i}+\tau_\mathrm{m})/e$) as a function of $\tau_\mathrm{m}$ at $\tau_\mathrm{i} = 0.2~\mu$s and $V_\mathrm{A} = 0.8~$V (see black dashed line in b). 
As expected, the $\ket{\uparrow}$-transport shows a linear dependency on $\tau_\mathrm{m}$, corresponding to a steady tunnel current (see schematic in a), whereas the $\ket{\downarrow}$-transport saturates due to an occupation of the ground state (see schematic in a). The solid line represents a fit according to \rev{$\langle n \rangle/\mathrm{pulse} = \Gamma_\text{D}(1-e^{-\gamma \tau_\mathrm{m}})/2\gamma$}.
}}
\label{f2}
\end{figure*}

The single particle spectrum of the QD can be resolved by finite bias spectroscopy measurements of the $ \mathcal{N}=0 \rightarrow 1$ electron transition (Fig.~1d). At finite magnetic field, the two energetically lower $K'$ valley polarized spin states as well as the nearly degenerate $K$-states can be well observed (see arrows and dashed lines in Fig.~\ref{f1}d). 
Fig.~\ref{f1}e shows the extracted splitting $\Delta E$ of the two spin states in the $K'$-valley, from now on denoted as $\ket{\downarrow}$ and $\ket{\uparrow}$, as a function of $B_\perp$. From the slope we determine $\Delta_\mathrm{SO} = 66\pm8~\mu$eV and \rev{$g_\text{s} = 1.93 \pm 0.09$}, which is in good agreement with earlier experiments~\cite{Banszerus2021Sep,Kurzmann2021Mar}.

\begin{figure*}[]
\centering
\includegraphics[draft=false,keepaspectratio=true,clip,width=\linewidth]{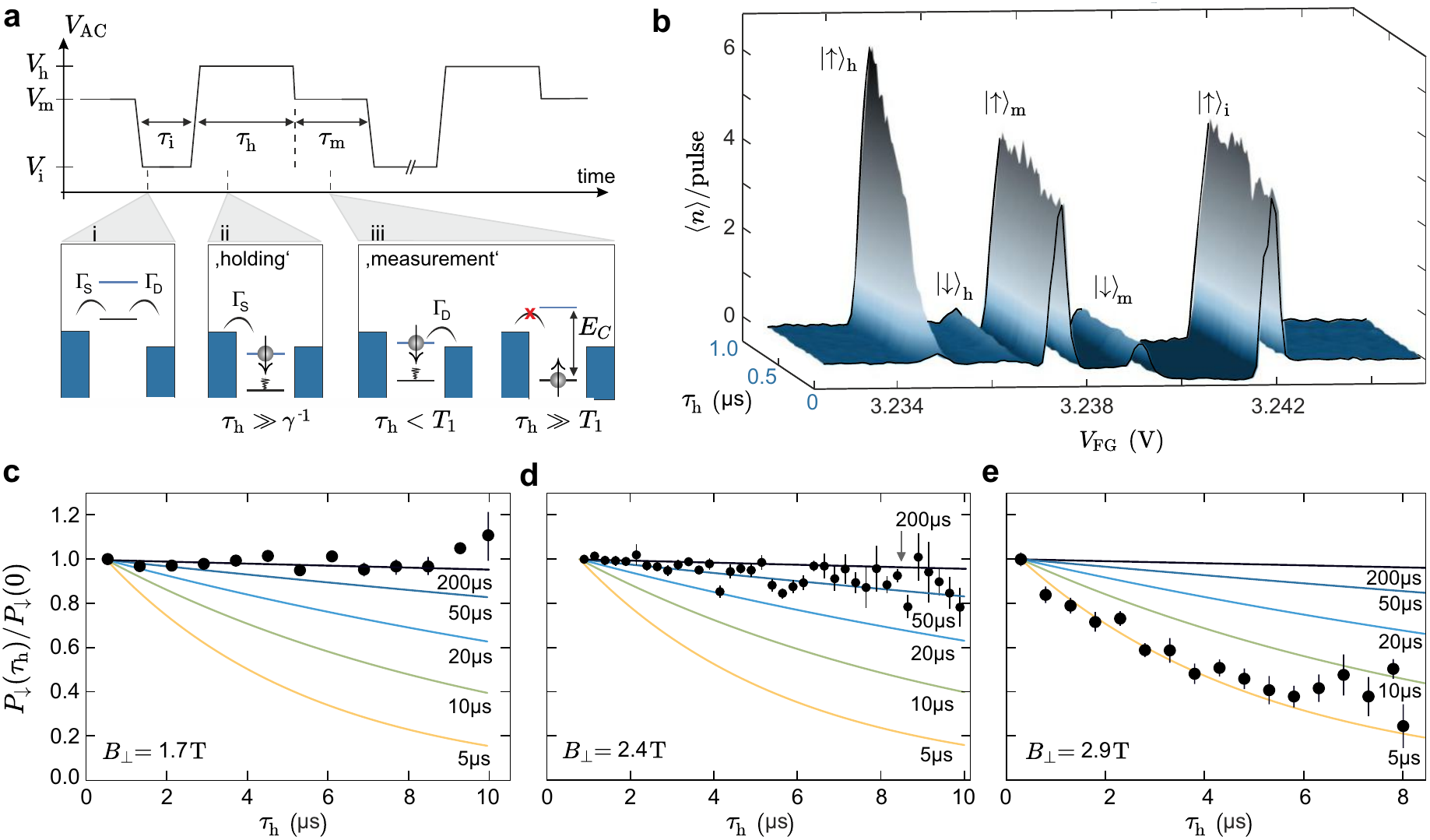}
\caption[Fig03]{\textbf{Measurement of the spin relaxation time.}
\textbf{a} Schematic of the applied three-level pulse train characterized by the voltages $V_\mathrm{i}$, $V_\mathrm{h}$, $V_\mathrm{m}$ and the times $\tau_\mathrm{i}$, $\tau_\mathrm{h}$, $\tau_\mathrm{m}$. During initialization ($\tau_\mathrm{i}$), the QD is emptied. Subsequently, both $\ket{\uparrow}$ and $\ket{\downarrow}$ are pushed below the bias window during $\tau_\mathrm{h}$ allowing tunneling from the reservoirs into either of the states. Furthermore, relaxation from $\ket{\downarrow}$ to $\ket{\uparrow}$ is possible. In the readout step ($\tau_\mathrm{m}$), $\ket{\downarrow}$ is aligned in the bias window, i.e. an electron in $\ket{\downarrow}$ can leave the QD contributing to the current. 
\textbf{b} Average number of electrons per pulse cycle $\langle n \rangle/\mathrm{pulse} = I(\tau_\mathrm{i}+\tau_\mathrm{h}+\tau_\mathrm{m})/e$ as a function of $V_\mathrm{FG}$ and $\tau_\mathrm{h}$ ($\tau_\mathrm{i} = 0.4~\mu$s, $\tau_\mathrm{m} = 0.4~\mu$s, $V_\mathrm{i} = -1~$V, $V_\mathrm{h} = 0.6~$V, $V_\mathrm{m} = 0~$V and $B_\perp = 2.4~$T).
\rev{Individual line cuts of the data set are shown in supplementary Fig.~S4.}
\textbf{c-e} The probability $P_\downarrow (\tau_\mathrm{h})/P_\downarrow (0)$ of the electron to remain in the excited state during $\tau_\mathrm{h}$ as a function of $\tau_\mathrm{h}$. Data has been acquired at $B_\perp=$1.7~T, 2.4~T and 2.9~T, respectively. Solid curves correspond to calculations considering different spin relaxation times $T_1$.
}
\label{f3}
\end{figure*}

To gain insights on the relaxation of the 
$\ket{\downarrow}$ excited state to the $\ket{\uparrow}$ ground state, we now focus on transient current spectroscopy measurements.
First, we use a two-level pulse scheme~\cite{Fujisawa2001Feb,Hanson2003Nov,Volk2013Apr} to extract the combined tunneling and the overall blocking rate of the system. 
time in BLG QDs~\cite{Banszerus2021Feb}.
We therefore apply a finite magnetic field of $B_\mathrm{\perp}=2.4$~T to lift the spin and valley degeneracy and, furthermore, to reduce the tunneling rates to the reservoirs~\cite{Eich2018Jul,Banszerus2021Feb}, \rev{by altering the density of states in the reservoirs~\cite{Banszerus2020Dec} and widening the tunneling barriers.}
Fig.~\ref{f2}a shows the applied square pulse scheme with amplitude $V_\text{A}$ and pulse widths $\tau_\text{i}$ and $\tau_\text{m}$. During $\tau_\mathrm{i}$, the QD is emptied (initialized). 
If the ground state $\ket{\uparrow}$ is in the bias window ($eV_{\text{SD}}$) during $\tau_\mathrm{m}$, a steady current can be observed. If the excited state $\ket{\downarrow}$ is in the bias window during $\tau_\mathrm{m}$, a transient current can be present, where electrons tunnel through the QD until one relaxes with a spin-flip or the ground state $\ket{\uparrow}$ gets occupied by direct tunneling from the reservoir. 
The current, $I$, through the device as a function of the pulse amplitude $V_\mathrm{A}$ and $V_\mathrm{FG}$ is shown in Fig.~\ref{f2}b. The two dominant transitions originate from $\ket{\uparrow}$-transport during $\tau_\mathrm{i}$ and $\tau_\mathrm{m}$ (see white dashed lines in Fig.~\ref{f2}b and left schematic in Fig.~2a). If the pulse amplitude %at the sample
exceeds the energy splitting of $\ket{\uparrow}$ and $\ket{\downarrow}$, a transient current can be observed during $\tau_\mathrm{m}$ (see black dashed line in Fig.~\ref{f2}c and right schematic in Fig.~3a). Importantly, the rise time of the pulses \rev{needs} to be faster than the inverse tunneling rates, such that the system cannot follow the pulse adiabatically \rev{(see supplementary Fig.~S3 for details)}.

Studying the dependence of the transient current on the pulse width $\tau_m$, we can extract quantitative information on the characteristic time scales of transient processes. 
Fig.~\ref{f2}c shows the average number $\langle n \rangle$ of electrons tunneling per pulse cycle.
As expected, in case of $\ket{\uparrow}$-transport, $\langle n \rangle$/pulse increases linearly with $\tau_\mathrm{m}$, where the slope is given by the combined tunneling rate of both barriers $\Gamma = \Gamma_\text{S}\Gamma_\text{D}/(\Gamma_\text{S}+\Gamma_\text{D}) \approx 6.6$~MHz. 
Transport via $\ket{\downarrow}$ saturates, as the probability of blocking transport by relaxation or tunneling from the reservoir increases with $\tau_\mathrm{m}$. 
To enhance transient currents, we establish an asymmetry between the source and the drain tunneling rate $\Gamma_\text{S} \gg \Gamma_\text{D}$ by tuning a FG adjacent to the QD.
Assuming spin-independent tunnel rates, in this regime, the number of electrons tunneling via $\ket{\downarrow}$ can be approximated by \rev{$\langle n \rangle/\mathrm{pulse} = \Gamma_\text{D}(1-e^{-\gamma \tau_\mathrm{m}})/2\gamma$}, with the blocking rate $\gamma$~\cite{Hanson2003Nov}.  
A fit of the data yields a blocking rate $\gamma \approx 7.9~$MHz and $\Gamma_\text{D} \approx 6.6~$MHz.
As $\gamma$ is on the order of $\Gamma_\text{D}$, direct tunneling from the reservoir into $\ket{\uparrow}$ dominates the blocking rate and, hence, the blocking rate only provides a lower bound for the relaxation time $T_1$. 

To extract $T_1$, we then follow Refs.~\cite{Fujisawa2002Sep,Hanson2003Nov} and include an additional voltage step in the pulse scheme, which allows separating the 'relaxation' from the 'measurement' step. The corresponding three-level pulse scheme is  depicted in Fig.~3a, where the pulse segments are described by pulse durations ($\tau_\text{i}$, $\tau_\text{h}$ and $\tau_\text{m}$) and corresponding voltage values $V_\text{AC}=$ $V_\text{i}$, $V_\text{h}$ and $V_\text{m}$. 
During the initialization step $\tau_\mathrm{i}$, the QD is emptied (see \rev{schematic i} in Fig.~3a). 
Next, both states $\ket{\uparrow}$ and $\ket{\downarrow}$ are pushed below the bias window in the loading and holding step ($\tau_\mathrm{h}$, $V_\mathrm{h}$). If $\tau_\text{h} \gg \gamma^{-1}$, it is ensured that an electron has tunneled into either one of the two states (see \rev{schematic ii} in Fig.~\ref{f3}a). Finally, to allow for spin-selective readout during the measurement step ($\tau_\mathrm{m}$, $V_\mathrm{m}$), the QD levels are aligned such that only an $\ket{\downarrow}$-electron (i.e. an electron that has not relaxed) can tunnel out to the drain and contribute to the current (see \rev{schematic iii} of Fig.~\ref{f3}a). 
Fig.~\ref{f3}b shows $\langle n \rangle/$pulse as a function of $V_{\text{FG}}$ and $\tau_\mathrm{h}$. The three transitions labeled $\ket{\uparrow}_\mathrm{i,h,m}$ originate from $\ket{\uparrow}$ ground state transport during ($\tau_\mathrm{i}$,$V_\mathrm{i}$), ($\tau_\mathrm{h}$,$V_\mathrm{h}$) and ($\tau_\mathrm{m}$,$V_\mathrm{m}$), respectively. As in Fig.~\ref{f2}c, the $\ket{\uparrow}_\mathrm{h}$ amplitude increases linearly with the duration the ground state is in the bias window, while $\ket{\downarrow}_\mathrm{h}$ saturates with the characteristic blocking rate, $\gamma$, of the system. The peak labeled $\ket{\downarrow}_\mathrm{m}$ originates from the electrons leaving the $\ket{\downarrow}$ excited state to the drain during the measurement step. The slight negative background between $\ket{\downarrow}_\mathrm{m}$ and $\ket{\uparrow}_\mathrm{i}$, stems from statistical backwards pumping of electrons during $\tau_\text{i}$.  
The relaxation time, $T_1$, can be determined from the amplitude of the $\ket{\downarrow}_\mathrm{m}$-peak. In order to contribute to $\ket{\downarrow}_\mathrm{m}$, electrons have to remain in the excited state and not relax during $\tau_\mathrm{h}$. The amplitude of $\ket{\downarrow}_\mathrm{m}$ as function of $\tau_\mathrm{h}$ is directly proportional to the probability $P_\downarrow(\tau_\mathrm{h})$ of an electron remaining in the excited state during $\tau_\mathrm{h}$. Figs.~\ref{f3}c-e show data sets for different $B_\perp$ which have been normalized according to $\langle n (\tau_\mathrm{h})\rangle /\langle n (0)\rangle = P_\downarrow(\tau_\mathrm{h})/P_\downarrow(0) = e^{-\tau_\mathrm{h}/T_1}$ following Ref.~\cite{Hanson2003Nov}. Hence, the data is expected to follow an exponential decay, where $T_1$ is the decay constant.
The solid lines in Figs.~\ref{f3}c-e show the exponential decay of $P_\downarrow(\tau_\mathrm{h})/P_\downarrow(0)$ for different values of $T_1$. At $B_\perp=1.7$~T (Fig.~\ref{f3}c), no decay of $P_\downarrow(\tau_\mathrm{h})/P_\downarrow(0)$ as function of $\tau_\mathrm{h}$ can be observed within the noise level of the data and a lower bound of $T_1 > 200~\mu$s is estimated from the comparison of the data with the calculated traces. At higher magnetic fields, i.e. $B_\perp=2.4$~T (see Fig.~\ref{f3}d) a slight and almost linear decay of $P_\downarrow(\tau_\mathrm{h})/P_\downarrow(0)$ can be observed, which is compatible with $T_1 \approx 50~\mu$s. When further increasing the magnetic field to $B_\perp=2.9$~T a clear exponential decay of $P_\downarrow(\tau_\mathrm{h})/P_\downarrow(0)$ with $T_1 \approx 5~\mu$s can be observed (see Fig.~\ref{f3}e).

\begin{figure}[]
\centering
\includegraphics[draft=false,keepaspectratio=true,clip,width=\linewidth]{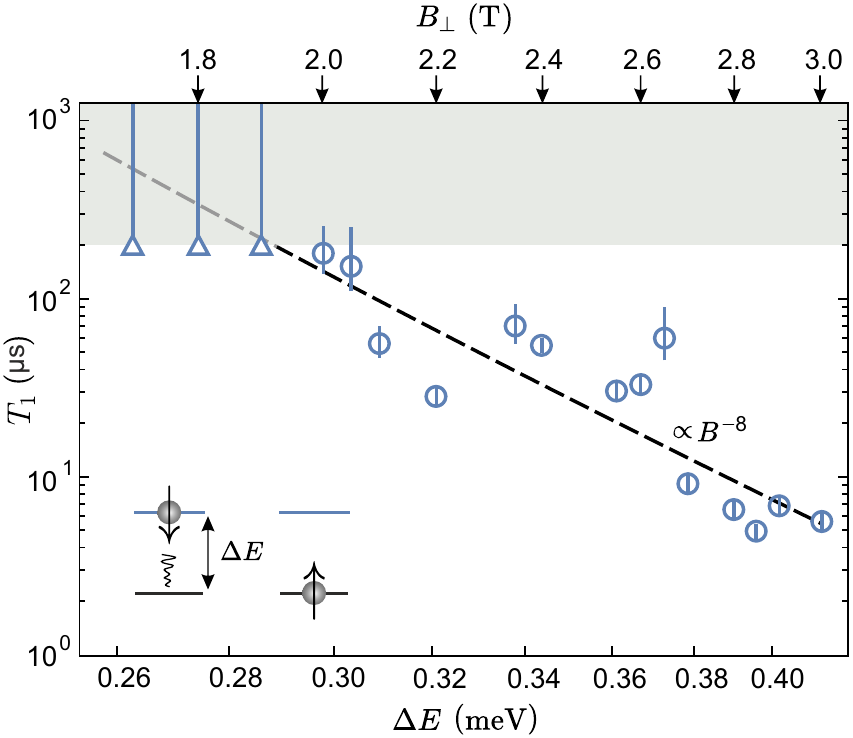}
\caption[Fig04]{\textbf{Dependence of $T_1$ on the spin splitting.} Spin relaxation time $T_1$ as a function of the spin splitting $\Delta E=~\Delta_\mathrm{SO}+g_s\mu_B B_\perp$ and the magnetic field, $B_\perp$, 
%(bottom axis) and the applied $B_\perp$ (top axis) 
on a double logarithmic scale. The gray shaded region marks the regime where only a lower bound for $T_1$ can be stated due to a limitation by the signal-to-noise ratio of the measurement. \rev{The dashed line marks a power law of $T_1 \propto B^{-8}$}. The error bars indicate the 1$\sigma$ confidence interval of an exponential fit to the data. 
}
\label{f4}
\end{figure}

Fig.~\ref{f4} shows $T_1$ \rev{times} extracted from exponential fits (round data points) to additional data sets as depicted in Figs.~\ref{f3}c-e as a function of the energy splitting $\Delta E$ and, hence, $B_\perp$ \rev{(see arrows)}.
Decreasing the magnetic field from $B_\perp =$ 3 to 2~T, $T_1$ increases by almost two orders of magnitude from about 5~$\mu$s to  200~$\mu$s. For magnetic fields below $B_\perp =2$~T, no exponential decay of $P_\downarrow(\tau_\mathrm{h})/P_\downarrow(0)$ can be fitted to the data anymore and only a lower bound of $T_1 > 200~\mu$s, can be stated (see triangular data points), limited by the signal-to-noise ratio of the measured data. \rev{Upon increasing $\tau_\text{h}$ (during which no current tunnels through the QD), the average current and thus the measurement signal decreases, limiting $\tau_\text{h}$ to 10~$\mu$s, before the signal-to-noise ratio decreases below one.}\\ 
\rev{Although our $B_\perp$-field range is limited, the strong dependence of the extracted $T_1$ times as function of the magnetic field (best described by a power law of $T_1 \propto  B^{-8}$,  
see dashed line in Fig.~4) may provide important insights on the spin relaxation mechanism. From detailed (theoretical) studies of the $B$-field dependent $T_1$ times in GaAs QDs~\cite{Golovach2004Jun,Hanson2007Oct,Camenzind2018Aug}, Si QDs~\cite{Raith2011May,Watson2017Mar,Hollmann2020Mar} and single-layer graphene nanoribbon-based QDs~\cite{Droth2013May} it is known that the spin-orbit coupling and the electron-phonon (e-ph) coupling, in particular the coupling of piezoelectric or acoustic phonons to electrons~\cite{Hanson2007Oct} are playing a crucial role for relaxation.
Indeed, it has been shown that a power-law decrease of $T_1$ as function of increasing spin splitting $\Delta E \propto B$~\footnote{\rev{The spin splitting $\Delta E$ is composed of the Zeeman splitting, which increases linearly with $B$, as well as the constant 'Zeeman-like' Kane-Mele spin orbit gap, $\Delta_\mathrm{SO}$ (c.f. Figure 1e). 
Thus for graphene and BLG $\Delta E \propto B$ is strictly speaking only valid if one neglects $\Delta_\mathrm{SO}$.}} originates in such systems from enhanced phonon emission due to both, an increasing phonon density of state and an increasing (accoustic) phonon momentum with increasing $\Delta E$, which in turn leads to faster spin relaxation for larger $B$-fields~\cite{Hanson2007Oct}.
The exact exponent of the power law scaling depends, however, sensitively on the system specific nature of (i) the e-ph coupling mechanisms, (ii) the phonons involved, (iii) the spin-orbit coupling, as well as (iv) the overall dimensionality of the system.
For example, for GaAs QDs a $T_1 \propto B^{-5}$ power law has been reported for $B$-fields in the range of $2 - 6$~T~\cite{Hanson2007Oct,Camenzind2018Aug}, while for small $B$-fields and suppressed spin-orbit coupling also a $T_1 \propto B^{-3}$ dependence has been observed~\cite{Camenzind2018Aug}.
Interestingly, for Si QDs a significantly stronger power-law, $T_1 \propto B^{-7}$, has been predicted and observed for $B > 2$~T~\cite{Raith2011May,Hollmann2020Mar}, which for multidonor
QDs in Si is reduced to a $T_1 \propto B^{-5}$ scaling, hlighting the sensitive dependence on microscopic details. 
While for single-layer graphene armchair nanoribbon-based QDs an e-ph coupling dominated $T_1 \propto B^{-5}$ is theoretically predicted for $B<3$~T there is -- to the best of our knowledge -- no theory yet for electrostatically confined QDs in BLG.
As the e-ph coupling in single-layer graphene nanoribbons and BLG are fundamentally different (just to mention the different dimensionality and the  dominant gauge-field coupling in single-layer graphene~\cite{Sohier2014Sep}) it is very hard to make at the present stage any prediction of what the theoretically expected power-law dependence for BLG QDs should be.
With almost certainty, electron-phonon coupling will also play an important role for BLG QDs and the observed strong $B$-field dependence of the $T_1$ time, which  gives hope for even longer times at smaller $B$-fields, may also point to a modified BLG phonon bandstructure when encapsulated in hBN. We expect that our experimental observation will trigger dedicated theoretical work on the spin relaxation in BLG QDs.
}

It is important to mention, that our extracted $T_1$ times can be considered as sufficiently long for single-electron spin manipulation and mark an important step towards the implementation of
spin qubits in graphene.
Interestingly, the reported $T_1$ times are more than two orders of magnitude larger than the values reported for carbon nanotubes in a similar magnetic field range~\cite{Churchill2009Apr}, most likely thanks to the smaller spin-orbit interaction in BLG. 
To investigate \rev{ $T_1$ times at smaller spin splittings, where spin qubits could be operated}, the fabrication of devices with sufficiently opaque tunneling barriers is required, in order to achieve low tunneling rates at lower magnetic fields. Additionally, integrated charge sensors will be needed to allow for single-shot charge and spin detection. \\

\textbf{Methods}\\
The device was fabricated from a BLG flake encapsulated between two hBN crystals of approximately 25~nm thickness using conventional van-der-Waals stacking techniques. A graphite flake is used as a BG. Cr/Au SGs with a lateral separation of 80~nm are deposited on top of the heterostructure. Isolated from the SGs by 15~nm thick atomic layer deposited Al$_2$O$_3$, we fabricate 70~nm wide FGs with a pitch of 150~nm. For details of the fabrication process, we refer to Ref.~\cite{Banszerus2020Oct}.

In order to perform pulsed-gate experiments, the sample is mounted on a custom-made printed circuit board. The DC lines are low-pass-filtered (10~nF capacitors to ground). All FGs are connected to on-board bias-tees, allowing for AC and DC control on the same gate. The AC lines are equipped with cryogenic attenuators of -26~dB. $V_\text{AC}$ refers to the AC voltage applied prior to attenuation. All measurements are performed in a $^3$He/$^4$He dilution refrigerator at a base temperature of around 10~mK and at an electron temperature of around 60~mK using standard DC measurement techniques.
Throughout the experiment, a constant BG voltage of $V_\mathrm{BG} = -3.5~$V and a SG voltage of $V_\mathrm{SG} = 1.85~$V is applied to define a p-type channel between source and drain.\\

\textbf{Acknowledgements}\\
The authors thank G. Burkard, A. Hosseinkhani \rev{and L. Schreiber} for fruitful discussions, F. Lentz, S. Trellenkamp and D. Neumeier for help with sample fabrication and J. Klos for help with the SEM micrographs. This project has received funding from the European Union's Horizon 2020 research and innovation programme under grant agreement No. 881603 (Graphene Flagship) and from the European Research Council (ERC) under grant agreement No. 820254, the Deutsche Forschungsgemeinschaft (DFG, German Research Foundation) under Germany's Excellence Strategy - Cluster of Excellence Matter and Light for Quantum Computing (ML4Q) EXC 2004/1 - 390534769, through DFG (STA 1146/11-1), and by the Helmholtz Nano Facility~\cite{Albrecht2017May}. K.W. and T.T. acknowledge support from the Elemental Strategy Initiative conducted by the MEXT, Japan (Grant Number JPMXP0112101001) and  JSPS KAKENHI (Grant Numbers 19H05790, 20H00354 and 21H05233).\\\\
\textbf{Data availability}\\
The data supporting the findings are available in a Zenodo repository under accession code XXX. \\\\
\textbf{Author contributions}\\
C.S. designed and directed the project; L.B., K.H., S.M. and E.I. fabricated the device, L.B., K.H. and C.V. performed the measurements and analyzed the data. K.W. and  T.T.  synthesized  the  hBN  crystals. C.V. and C.S. supervised the project. L.B., K.H., C.V. and C.S. wrote the manuscript with contributions from all authors. L.B. and K.H. contributed equally to this work. Correspondence should be addressed via e-mail to luca.banszerus@rwth-aachen.de \\\\
\textbf{Competing interests}\\
The authors declare no competing interests.

\bibliography{literature}

\end{document}

\author{L.~Banszerus$^*$}
\email{luca.banszerus@rwth-aachen.de.}
\affiliation{JARA-FIT and 2nd Institute of Physics, RWTH Aachen University, 52074 Aachen, Germany,~EU}%
\affiliation{Peter Gr\"unberg Institute  (PGI-9), Forschungszentrum J\"ulich, 52425 J\"ulich,~Germany,~EU}

\author{K.~Hecker$^*$}
\affiliation{JARA-FIT and 2nd Institute of Physics, RWTH Aachen University, 52074 Aachen, Germany,~EU}%
\affiliation{Peter Gr\"unberg Institute  (PGI-9), Forschungszentrum J\"ulich, 52425 J\"ulich,~Germany,~EU}

\author{S.~M\"oller}
\affiliation{JARA-FIT and 2nd Institute of Physics, RWTH Aachen University, 52074 Aachen, Germany,~EU}%
\affiliation{Peter Gr\"unberg Institute  (PGI-9), Forschungszentrum J\"ulich, 52425 J\"ulich,~Germany,~EU}

\author{E. Icking}
\affiliation{JARA-FIT and 2nd Institute of Physics, RWTH Aachen University, 52074 Aachen, Germany,~EU}%
\affiliation{Peter Gr\"unberg Institute  (PGI-9), Forschungszentrum J\"ulich, 52425 J\"ulich,~Germany,~EU}

%\author{S.~Trellenkamp}
%\author{F.~Lentz}
%\affiliation{Helmholtz Nano Facility, Forschungszentrum J\"ulich, 52425 J\"ulich,~Germany,~EU}

%\author{D.~Neumaier}
%\affiliation{AMO GmbH, Gesellschaft f\"ur Angewandte Mikro- und Optoelektronik, 52074 Aachen, Germany, EU}

\author{K.~Watanabe}
\affiliation{Research Center for Functional Materials, 
National Institute for Materials Science, 1-1 Namiki, Tsukuba 305-0044, Japan
}
\author{T.~Taniguchi}
\affiliation{ 
International Center for Materials Nanoarchitectonics, 
National Institute for Materials Science,  1-1 Namiki, Tsukuba 305-0044, Japan
}

\author{C.~Volk}
\author{C.~Stampfer}
\affiliation{JARA-FIT and 2nd Institute of Physics, RWTH Aachen University, 52074 Aachen, Germany,~EU}%
\affiliation{Peter Gr\"unberg Institute  (PGI-9), Forschungszentrum J\"ulich, 52425 J\"ulich,~Germany,~EU}%

\title{Supplementary Information: \\ Spin relaxation in a single-electron graphene quantum dot}

\date{\today}% It is always \today, today,
\maketitle

%\section*{Supplementary Note 1: Details on sample design and tuning towards a single electron DQD} 

\subsection{Formation of a single-electron quantum dot}

\begin{figure}[b]
\includegraphics[draft=false,keepaspectratio=true,clip,width=1\linewidth]{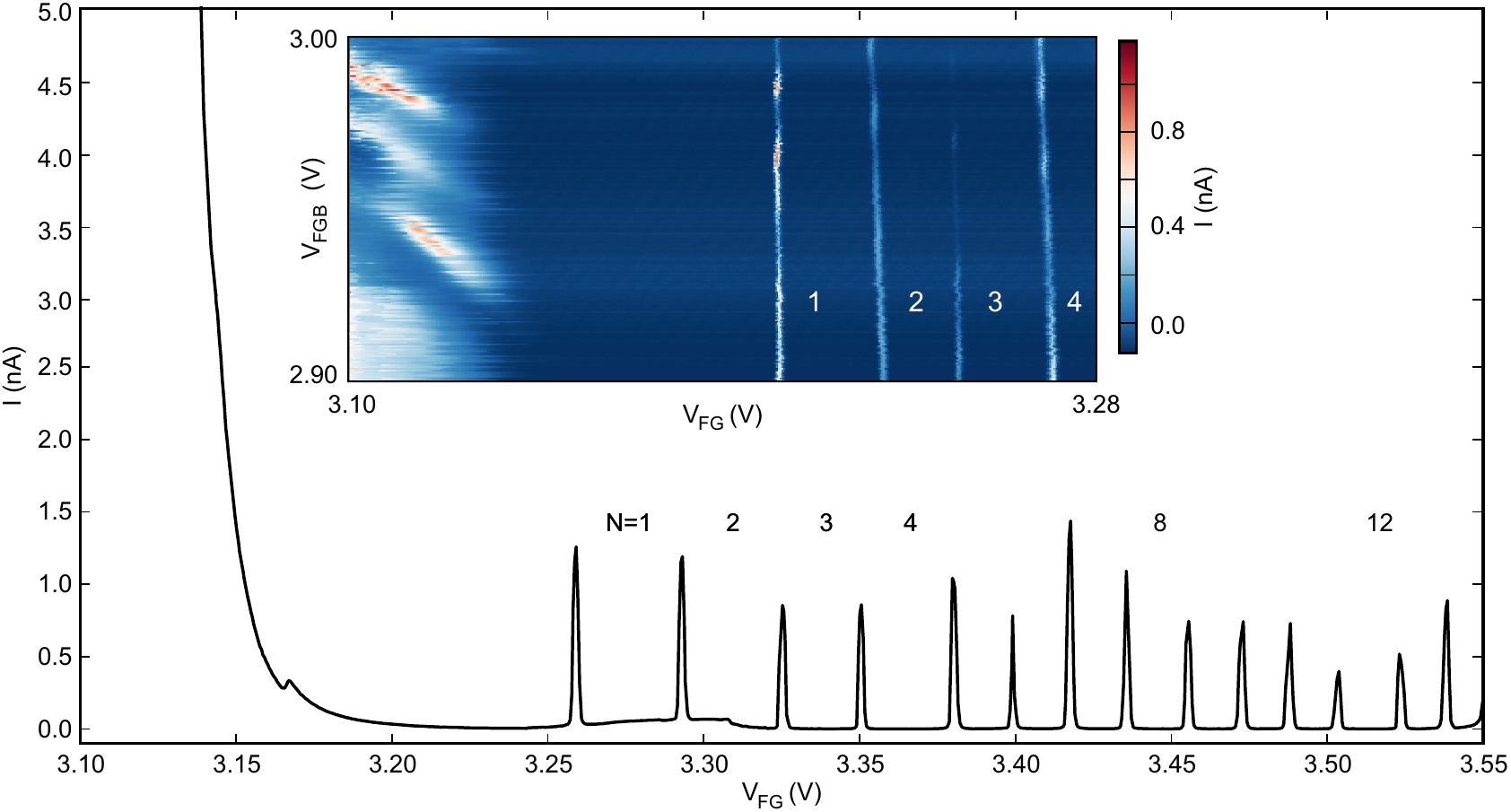}
\caption[FigS1]{Tunneling current as function of the finger gate potential $V_\mathrm{FG}$ ($V_\mathrm{SD} = 1$~mV). 
%The Coulomb resonances are grouped in quadruplets, reflecting the fourfold shell filling pattern. 
Numbers indicate the electron occupation of the QD. 
Inset: Tunneling current as function of $V_\mathrm{FG}$ and the voltage applied to the barrier gate $V_\mathrm{FGB}$. Note that there has been a charge rearrangement between the longer line trace and the measurement shown in the inset. 
}
\label{s1}
\end{figure}

A p-type channel is defined between the source and the drain by the voltages applied to the back gate and the split gates ($V_\mathrm{BG} = -3.5$~V, $V_\mathrm{SG} = 1.85$~V throughout the experiment). Fig.~\ref{s1} shows the current through the channel as a function of the voltage $V_\mathrm{FG}$ applied to the finger gate (FG) (see labeling in Figs.~1a,b of the main text).
Increasing $V_\mathrm{FG}$ locally compensates the potential set by the back gate and depletes the channel underneath the FG. Between $V_\mathrm{FG} \approx 3.2$~V and $V_\mathrm{FG} \approx 3.26$~V, the Fermi level lies in the band gap, hence the current is suppressed. 
A sequence of 14 Coulomb peaks can be observed starting from $V_\mathrm{FG} \approx 3.26$~V indicating the formation of a quantum dot as the conduction band is pushed below the Fermi level. 
The peak current through the QD remains almost constant as a function of the electron occupation. Furthermore, the Coulomb resonances are grouped in quadruplets (see labels 4, 8 and 12 in Fig.~S1) reflecting the fourfold shell-filling sequence, due to the spin and valley degeneracy in BLG. 
As the first four Coulomb resonances (as well as the next 2 $\times$ 4 peaks) are nicely grouped in four this indicates that they comprise a complete shell in agreement with the conclusion that we indeed see the 1st electron.

The inset in Fig.~\ref{s1} shows a charge stability diagramm, i.e. the current as function of $V_\mathrm{FG}$ and the barrier gate voltage, $V_\mathrm{FGB}$ (see green gate in Figs.~1a,b), highlighting that the Coulomb peaks are well visible over a larger barrier gate range.
With increasing $V_\mathrm{FGB}$, the conductance decreases and the peaks shift towards lower $V_\mathrm{FG}$ due to cross capacitance effects. Please note that no additional Coulomb peak corresponding to the QD can be observed at a lower voltage than the one labeled '1', independent on $V_\mathrm{FGB}$.

\begin{figure}[b]
\includegraphics[draft=false,keepaspectratio=true,clip,width=1\linewidth]{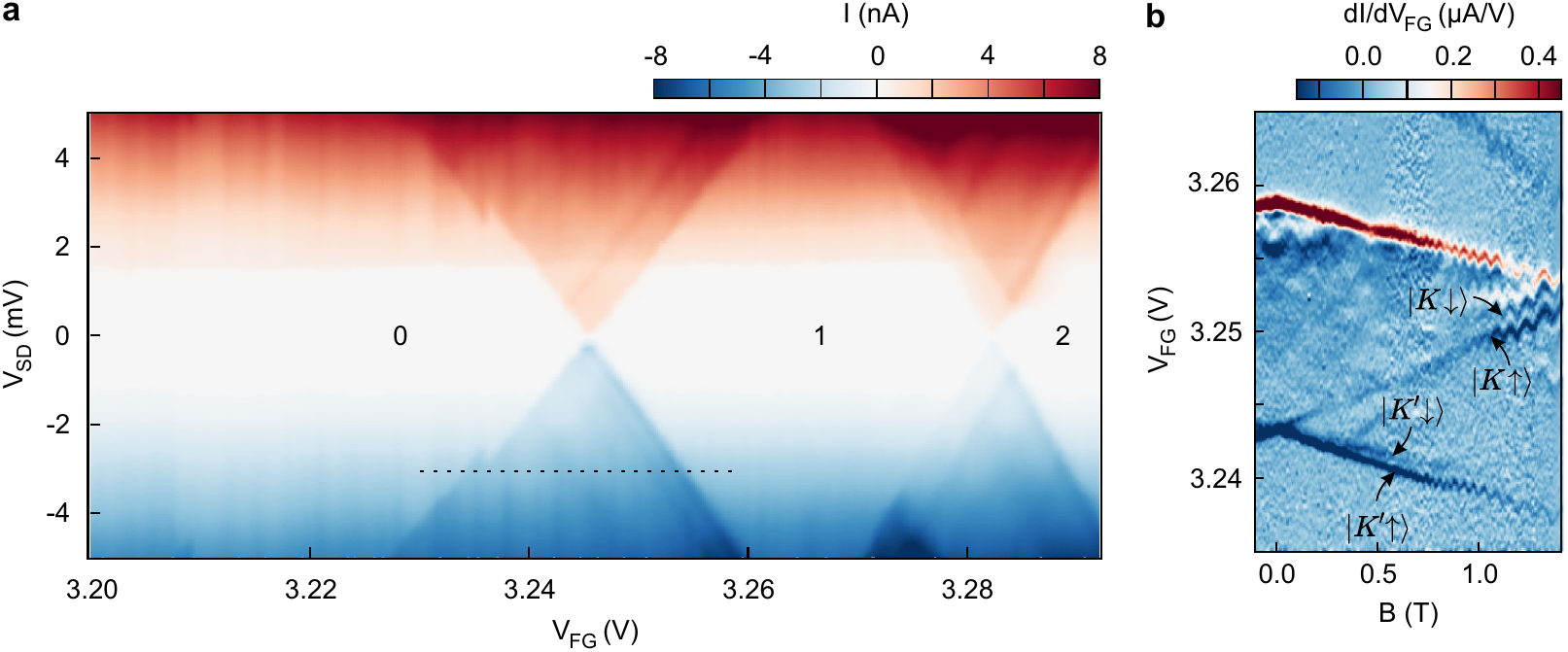}
\caption[FigS2]{ \textbf{a} Tunneling current as function of the finger gate voltage, $V_\mathrm{FG}$, and bias voltage, $V_\mathrm{SD}$, at a magnetic field of $B_\perp = 0$~T. Diamond shaped region of Coulomb blockade by occupation of the first electron can be observed. 
\textbf{b} Transconductance measured along the dashed line in \textbf{a} as function of the perpendicular magnetic field. (Note that the absolute position of the Coulomb peak has slightly changed between measurements due to charge rearrangements in the device.) Excited states of the first electron can be observed as resonant lines in the measurement, shifting in energy due to the spin and valley Zeeman effects. Note that the oscillations at higher $B$-fields are due to Shubnikov de Haas oscillations in the leads (see Banszerus et al. Phys. Status Solidi B 257, 2000333 (2020)).}
\label{s2}
\end{figure}

Fig.~\ref{s2}a shows a finite bias spectroscopy measurement around the first two Coulomb peaks at $B_\perp = 0$~T. 
At the addition of the first electron, no step in the current originating from an excited state can be observed (please compare to the 1-2 transition, where bias-symmetric current steps can be observed). Peaks in the current, which are asymmetric in bias voltage, arise from resonances within the density of the states in the leads and do not depend on the excited state spectrum of the QD.
Fig.~\ref{s2}b shows the transconductance along the dashed line in a as a function of $B=B_\perp$. Four states of the single-particle spectrum can be identified (see labels) which shift according to their spin and valley Zeeman effect. The measured spectrum is in perfect qualitative and quantitative agreement with the expected single-particle spectrum of a BLG QD which underlines that this transition corresponds indeed to the first electron of the QD.

\subsection{Bandwidth of the RF coaxial lines and the pulse generator}
\begin{figure}[h!]
\includegraphics[draft=false,keepaspectratio=true,clip,width=1\linewidth]{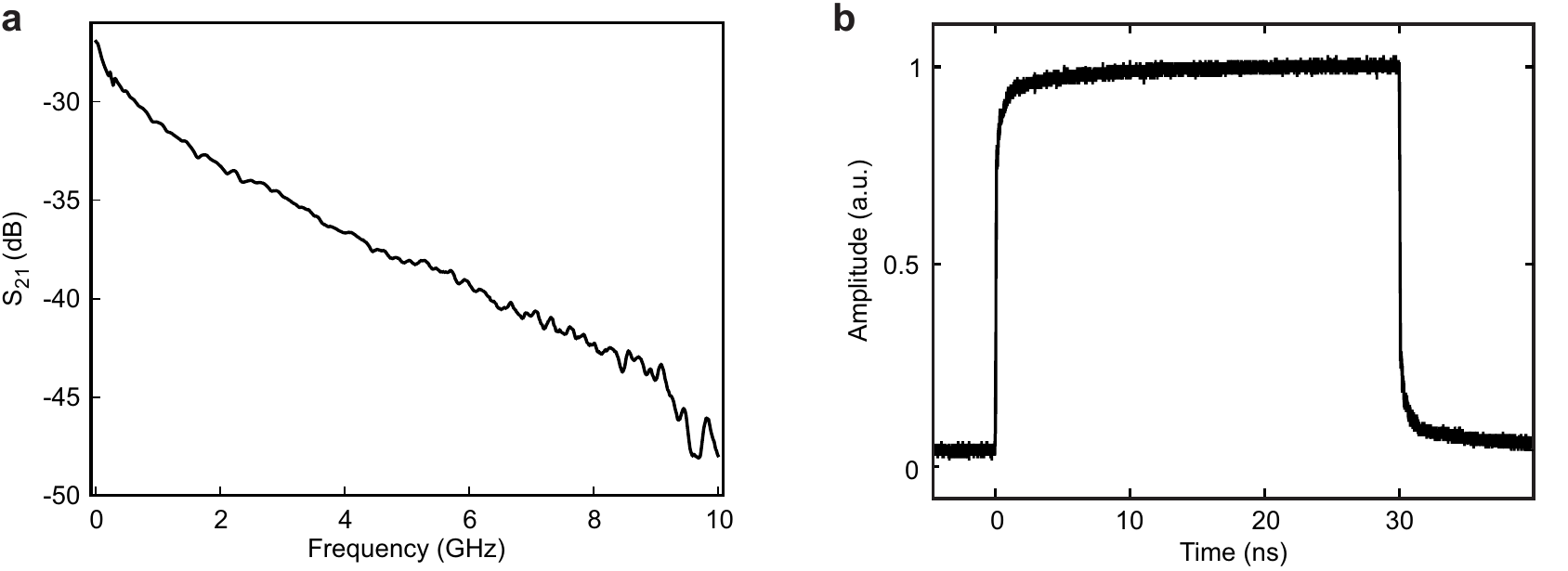}
\caption[FigS3]{\textbf{a} Transmission $S_{21}$ of a RF coaxial line in the cryostate as function of frequency. Attenuators with a total attenuation of -26~dB are installed in the cryostate. 
\textbf{b} Measured square pulse generated by the AWG. The rise time of the pulse is on the order of 100~ps.
%, much faster than the tunneling rates of the device.   
}
\label{s3}
\end{figure}

Fig.~\ref{s3}a shows the transmission $S_{21}$ of the coaxial line in the cryostate as function of frequency measured at room temperature.
Fig.~\ref{s3}b shows an exemplary square pulse generated by the arbitrary waveform generator used in the experiment (Tektronix AWG7082C). The rise time measures $\approx 100$~ps which is significantly shorter than the inverse of the tunneling rates ($\Gamma_\mathrm{S}, \Gamma_\mathrm{D}$ in the low MHz regime in our transient current spectroscopy experiments) fulfilling the condition of non-adiabaticity.

\subsection{Complementary data to Fig.~3b}
\begin{figure}[h!]
\includegraphics[draft=false,keepaspectratio=true,clip,width=0.6\linewidth]{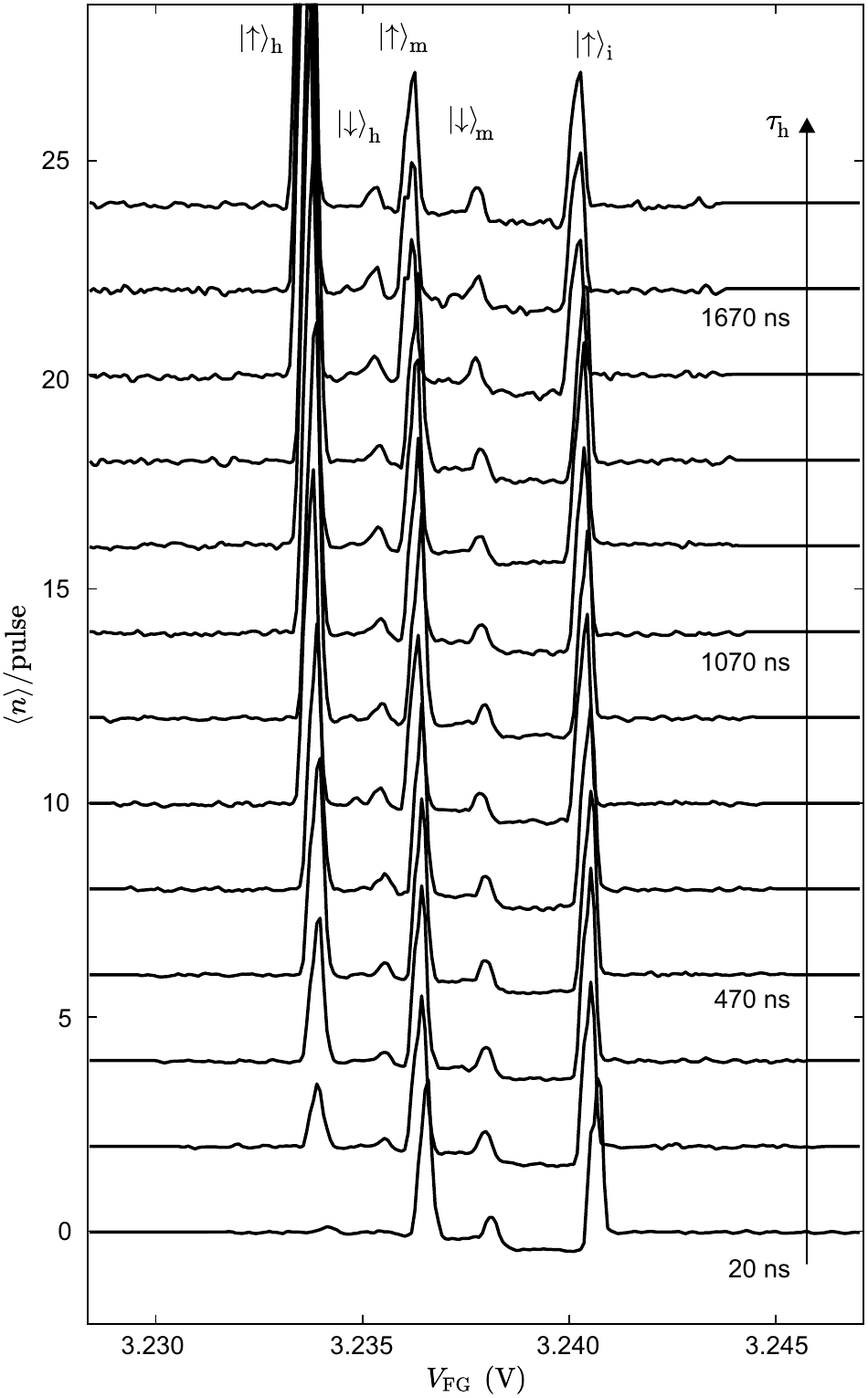}
\caption[FigS4]{Average number of electrons per pulse cycle $\langle n \rangle/\mathrm{pulse}$ as a function of $V_\mathrm{FG}$ for increasing $\tau_\mathrm{h}$ (in steps of 150~ns). The traces represent cuts through the data set of Fig.~3b of the main text. Traces are offset for clarity.
}
\label{s4}
\end{figure}

Fig.~\ref{s4} shows line cuts through the data set presented in Fig.~3b of the main text.
A small back pumping current is observed between $\ket{\downarrow}_m$ and $\ket{\uparrow}_i$ (around $V_\text{FG}=3.24$~V). Apart from that effect, the background signal is constant over the entire $V_\text{FG}$ regime. To extract the $\ket{\downarrow}_m$ amplitude and thus $P_\downarrow(\tau_\mathrm{h})$, we measure the peak height relative to the background on the left side of the peak. As the background level used as reference is on the same level as the background signal far away from the QD states, we conclude that no pumping effects take place.

%\newpage

%\bibliography{Literature}